\documentclass[FBSedit,FBSmath,ecsub]{FBSsuppl}
\usepackage{amsfonts}
\usepackage{amssymb}
\usepackage{epsfig}
\def\qed{\hbox{${\vcenter{\vbox{			
   \hrule height 0.4pt\hbox{\vrule width 0.4pt height 6pt
   \kern5pt\vrule width 0.4pt}\hrule height 0.4pt}}}$}}

\def\Journal#1#2#3#4{{#1} {\bf #2}, #3 (#4)}

\def\NPA{{\em Nucl. Phys.} A}
\def\PLB{{\em Phys. Lett.}  B}
\def\PRL{\em Phys. Rev. Lett.}
\def\PRC{{\em Phys. Rev.} C}

\def\ZPA{{\em Z. Phys.} A}

\def\EPJA{{\em Eur. Phys. J.} A}

\title{\hfill{\small {\bf MKPH-T-02-11}}\\
Three-Body Analysis of Incoherent Photoproduction of $\eta$ 
Mesons on the Deuteron near Threshold\thanks{Supported by 
Deutsche Forschungsgemeinschaft (SFB 443).}
} 
\author{A.\ Fix and H.\ Arenh\"ovel}
\institute{Institut f\"ur Kernphysik, 
Johannes Gutenberg-Universit\"at, 55099 Mainz, Germany}

\begin{document}

\maketitle
\vspace*{-.5cm}

\begin{abstract}
The importance of three-body dynamics in the $\eta np$ system in
elastic and inelastic $\eta$-deuteron scattering as well as coherent
and incoherent $\eta$ photoproduction on the deuteron in the 
energy region from threshold up to 30 MeV above has been investigated.
It is shown that a restriction to first order rescattering with
respect to the $NN$- and $\eta N$-final state interactions, i.e.,
restriction to rescattering in the two-body subsystems, does not give
a sufficiently accurate approximation to the $s$-wave reaction
amplitude and that higher order terms, as described by the three-body
dynamics give very substantial contributions.  
\end{abstract}

\section{Introduction}
Eta photoproduction on the deuteron is of interest with respect to
information on the neutron amplitude,
the role of $\eta N$ interaction;
With respect to the absolute strength of the neutron amplitude, 
quasi-free production in the incoherent process is favoured;
However, with respect to the relative phase between the elementary amplitudes 
of proton and neutron, the coherent reaction on the deuteron is 
better suited because the deuteron acts as an isospin filter and the 
proton and neutron amplitudes interfere coherently.

\section{The Elementary $\eta$ Production Operator}
\label{elem_operator}

\begin{figure}[hbt]
\vspace*{-.3cm}
\centerline{\epsfig{file=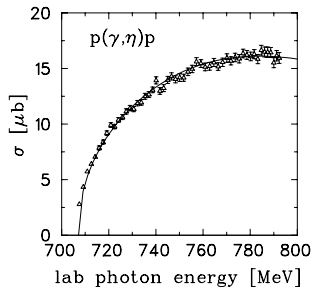,width=4cm}}
\vspace*{-.3cm}
\caption{Total cross section for $\eta$ production on the proton. Data
from Krusche {\it et al.}, \Journal{\PLB}{358}{40}{1995}.}
\label{fig1}
\vspace*{-.3cm}
\end{figure}

The following pure resonance form with excitation of the 
$S_{11}(1535)$ is used
\begin{eqnarray*}
t_{\gamma\eta}^{(s/v)}&=&\frac{k_{\gamma N}}{M_{N^*}+M_N}\,
\frac{e\,g_{\gamma NN^*}^{(s/v)}g_{\eta NN^*}}
{W_{\gamma N}-M_{N^*}+\frac{i}{2}
\Gamma(W_{\gamma N})}\,i\mbox{\boldmath$\sigma$}\cdot
\mbox{\boldmath$\varepsilon$}_{\lambda} \,.
\end{eqnarray*}
with parameters $M_{N^*}=1535$~MeV, $\Gamma_{\pi\pi N}=16$~MeV,
$g_{\eta NN^*}=2.10$, and $g_{\pi NN^*}=1.19$. 
The vertex constant $g_{\gamma NN^*}$ is related to  
the helicity amplitude $A_{1/2}^N$ by
$eg_{\gamma NN^*}=\sqrt{{2M_{N^*}(M_{N^*}\!+\!M_N)}/
{(M_{N^*}-M_N)}}A_{1/2}^N$.
This parametrization yields a good description of the total 
cross section on the proton using $A_{1/2}^p=0.104$ GeV$^{-1/2}$
as is shown in Fig.~\ref{fig1}.

\section{Inclusion of Final State Interaction in Incoherent $\eta$
Photoproduction by Rescattering in 2-Body Subsystems} 

\begin{figure}[h]
\centerline{\epsfig{file=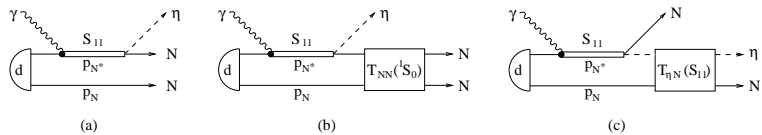,width=10cm}}
\vspace*{-.3cm}
\caption{Diagrams for $d(\gamma,\eta)np$: (a) IA, (b) $NN$
rescattering, (c) $\eta N$ rescattering.}
\label{fig2}
\vspace*{-.3cm}
\end{figure}

\begin{figure}[h] 
\vspace*{-.3cm}
\centerline{\epsfig{file=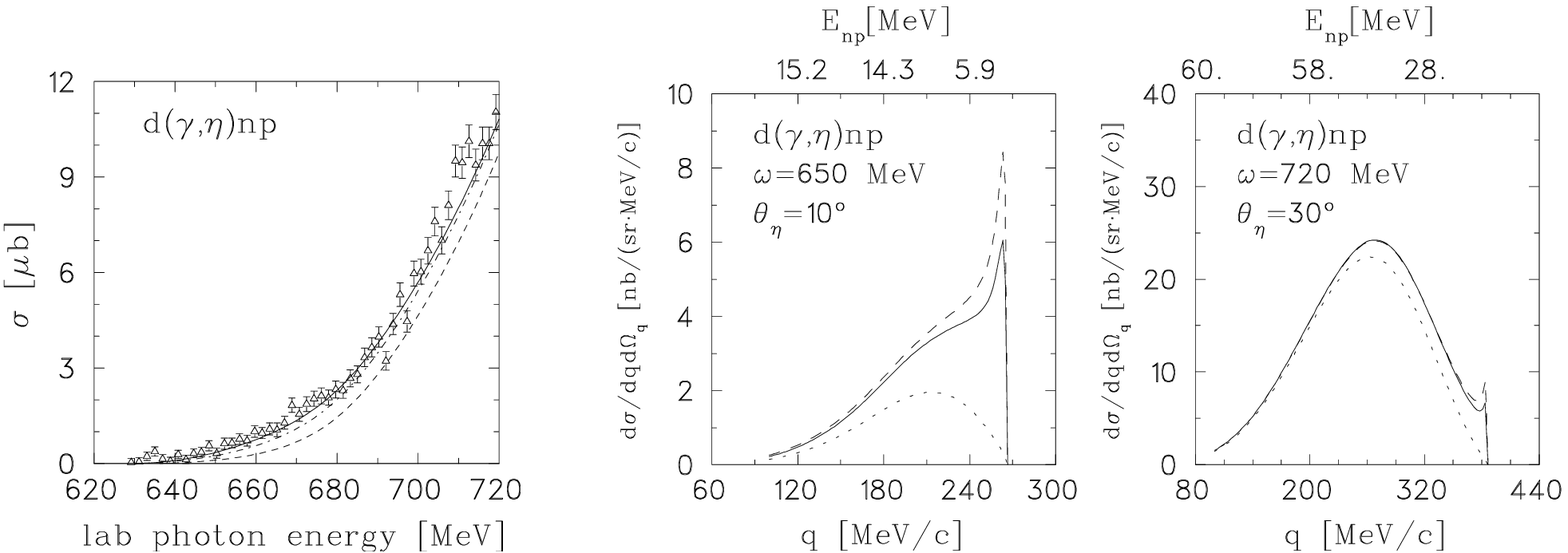,width=10cm}} 
\vspace*{-.3cm}
\caption{Left panel: Total cross section for $d(\gamma,\eta)np$:
dashed: IA, 
solid: IA + rescattering, dash-dotted: IA + $NN$ rescattering,
data: inclusive $\gamma d\!\rightarrow\eta X$ from Krusche {\it et al.}, 
\Journal{\PLB}{358}{40}{1995}.
Middel and right panels: $\eta$-meson spectra at forward angles: dotted: IA; 
solid: IA + rescattering; dashed: without $D$-wave contribution 
to $NN$-rescattering amplitude. Top abscissa indicates final
$NN$-excitation energy $E_{np}$. 
}
\label{fig3}
\vspace*{.1cm}
\centerline {\epsfig{file=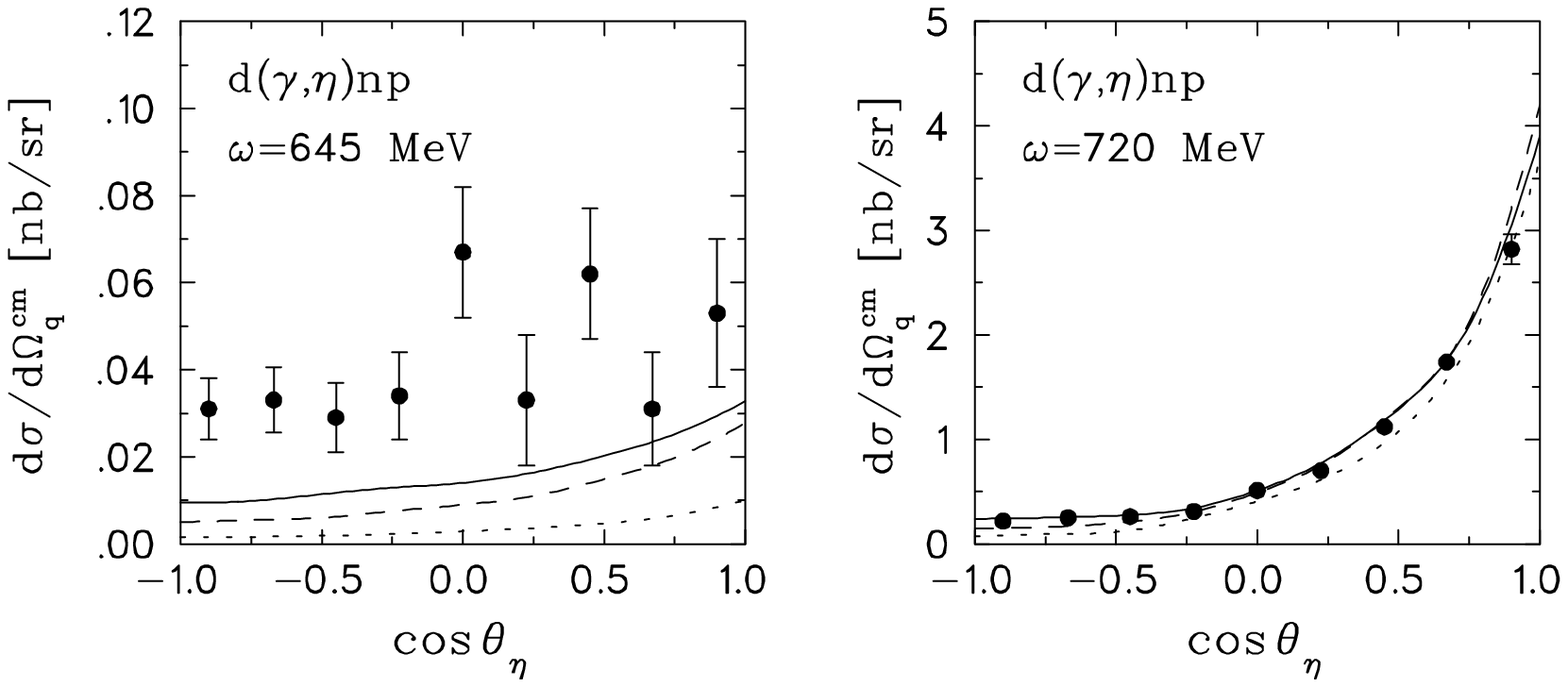,width=7cm}}
\vspace*{-.3cm}
\caption{Differential cross section for $d(\gamma,\eta)np$. 
Dotted: impulse approximation;
Dashed: IA plus $NN$ rescattering; Full: complete result. Exp.:
inclusive $\gamma d\!\rightarrow\eta X$ Krusche {\it et al.}, 
\Journal{\PLB}{358}{40}{1995}.} 
\label{fig4}
\vspace*{-.7cm}
\end{figure}

Near threshold the impulse approximation (IA) yields a very small 
cross section for $d(\gamma,\eta)np$ due to the
large momentum mismatch and indeed fails drastically yielding 
a cross section much too low compared to experiment. Therefore, we
first have performed an approximate treatment of final state
interaction (FSI) by taking into account only complete rescattering in
the two-body $\eta N$ and $NN$ subsystems of the final state (see
Fig.~\ref{fig2})~\cite{FiA97}. 
In this calculation, deuteron wave function and $T_{NN}$ are determined 
from the Bonn OBEPQ potential while for $\eta N$ rescattering 
the $\eta N$ $t$-matrix is taken in the isobar approach with 
intermediate excitation of the $S_{11}(1535)$ 
\begin{eqnarray*}
t_{\eta N}(W_{\eta N})&=&v_{N^*}^{\dagger}g_{N^*}(W_{\eta N})v_{N^*}
={g_{\eta NN^*}^2}{[W_{\eta N}-M_{N^*}+\frac{i}{2}\Gamma(W_{\eta N})]^{-1}}\,,
\end{eqnarray*}
In view of the near-threshold region it is sufficient to consider 
rescattering in $s$-waves only. This first order rescattering 
leads to a considerable improvement as is seen in the left panel of
Fig.~\ref{fig3}. 
The spectrum of the outgoing $\eta$ meson shows at low energies a 
distinct signature of the final state $NN$ rescattering exhibiting the
prominant $^1S_0$ peak close to threshold of $NN$-scattering 
(see middel and right panels of Fig.~\ref{fig3}).
Differential cross sections near threshold are shown in Fig.~\ref{fig4}
and one notes a considerable improvement over the IA. However, 
the left panel indicates that first order 
rescattering still fails to explain quantitatively the enhancement
of the experimental data right above threshold. 
This is corroborated by very recent more precise near-threshold data
by Hejny {\it et al.}, \Journal{\EPJA}{13}{493}{2002}.

\section{Three-Body Treatment of Final State Interaction in
$\eta$-Deuteron Scattering and $\eta$ Photoproduction} 

\begin{figure}
\centerline{\epsfig{file=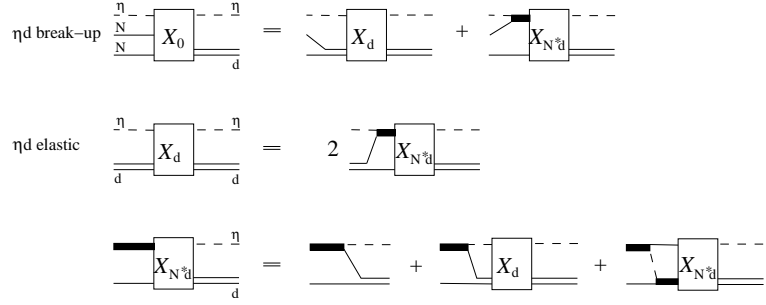,width=10cm}}
\vspace*{-.3cm}
\caption{Representation of three-body equations for
$\eta$-deuteron scattering.}
\label{fig5}
\vspace*{-.3cm}
\end{figure}

The very strong effect from the hadronic interaction in the final state 
in first order rescattering suggests that a genuine three-body treatment
is required. 
Defining as channels ``$N^*$'' the channel with one spectator nucleon
and ``$d$'' the channel with the $\eta$ meson as spectator and taking 
the interactions in separable form, one
obtains from the AGS-3-body equations a set of coupled equations for
the channel transition amplitudes (see Fig.~\ref{fig5} for a
diagrammatic representation), which reads
\begin{eqnarray*}
X_{N^*d} &=& Z_{N^*d}^{(\eta)}
+Z_{N^*d}^{(\eta)}\tau_d^{(\eta)}X_{d}^{(\eta)}
+Z_{N^*d}^{(\pi)}\tau_d^{(\pi)}X_{d}^{(\pi)}\\
&&+(Z_{N^*N^*}^{(\eta)}+Z_{N^*N^*}^{(\pi)})\tau_{N^*}X_{N^*d}\,,\\
X_{d}^{(\eta)}& =& 2Z_{dN^*}^{(\eta)}\tau_{N^*}X_{N^*d}\,, \quad
X_{d}^{(\pi)} = 2Z_{dN^*}^{(\pi)}\tau_{N^*}X_{N^*d}\,,
\end{eqnarray*}
where the $N^*$ channel comprises two components $N^*(\eta)$ and
$N^*(\pi)$ in order to account for the coupling of the $\eta N$-channel 
to the $\pi N$-channel. 
Because of the near threshold
region only $s$-waves are included. 
For the $NN$ interaction the tensor force is
neglected for reasons of simplicity. In detail, we have taken for 
the $NN$-channel a driving term $V_d(p,p')\,=\,g_d(p)\,g_d(p')$ 
with $g_d(p)={g_d\,\beta^2_d}/{(p^2+\beta^2_d)}$
yielding a $t$-matrix of the form 
$t_d(p,p')\,=\,g_d(p)\,\tau_d(E)\,g_d(p')$ with
\begin{eqnarray*}
\tau_d(E)\,=-\frac{1}{2M_N}\,
\Big[1+\frac{g_d^2\beta^2_d}{16\pi(\sqrt{EM_N}+i\beta_d)}\Big]^{-1} 
\end{eqnarray*}
Correspondingly, the $t$-matrix for the $\eta N$-channel has the form
$t^{(ij)}_{N^*}(p,p')\,=
\,g^{(i)}_{N^*}(p)\,\tau_{N^*}(E)\,g^{(j)}_{N^*}(p')$, where
$g^{(i)}_{N^*}(p)\,=\,
{g_{N^*}\,\beta^{(i)\,2}_{N^*}}/{(p^2+\beta^{(i)\,2}_{N^*})}$ and 
\begin{eqnarray*}
\tau_{N^*}(W)\,&=&\,[W-M_0-\Sigma_\pi(W)-\Sigma_\eta(W)]^{-1}\,.  
\end{eqnarray*}
For the values of the actual parameters we refer to~\cite{FiA02}. This 
interaction yields a scattering length $a_{\eta N}=0.75+i\,0.27$~fm.
\begin{figure}[h]
\centerline{\epsfig{file=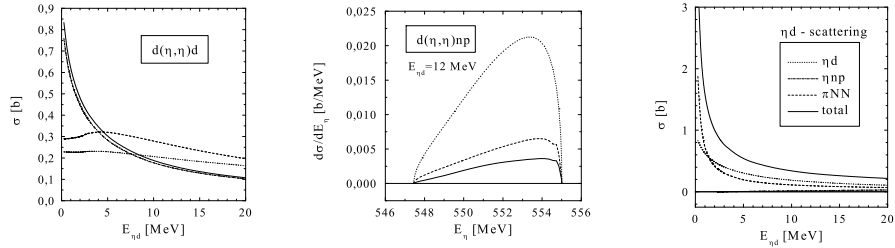,width=12cm}}
\vspace*{-.3cm}
\caption{Results for $\eta$-deuteron scattering
from~\protect\cite{FiA02}. Left panel: elastic total cross 
section, middle panel: spectrum of emitted $\eta$ mesons in inelastic
scattering, notation for left and middel panels: dotted: IA; dashed:
first order rescattering; solid: three-body calculation. Right panel:
various contributions to total $\eta d$ cross section.}
\label{fig6}
\vspace*{-.3cm}
\end{figure}
Results for $\eta d$ scattering are displayd in Fig.~\ref{fig6}.
For elastic scattering one notes a rapid increase of the total cross
section approaching the threshold (see left panel of Fig.~\ref{fig6}),
which is explained by the presence of a virtual pole in the $\eta NN$ 
system~\cite{FiA00}. This feature is not described by first order
rescattering. 
A similar failure is also seen for the inelastic channel (middel panel
of Fig.~\ref{fig6}). All contributions to the total cross section are
shown in the right panel of Fig.~\ref{fig6}. 

Turning now to $\eta$ photoproduction, an analogous representation as
in Fig.~\ref{fig5} holds for the coherent and incoherent
photoproduction amplitudes, replacing the incoming $\eta$ by a photon
line. 
\begin{figure}[h]
\centerline{\epsfig{file=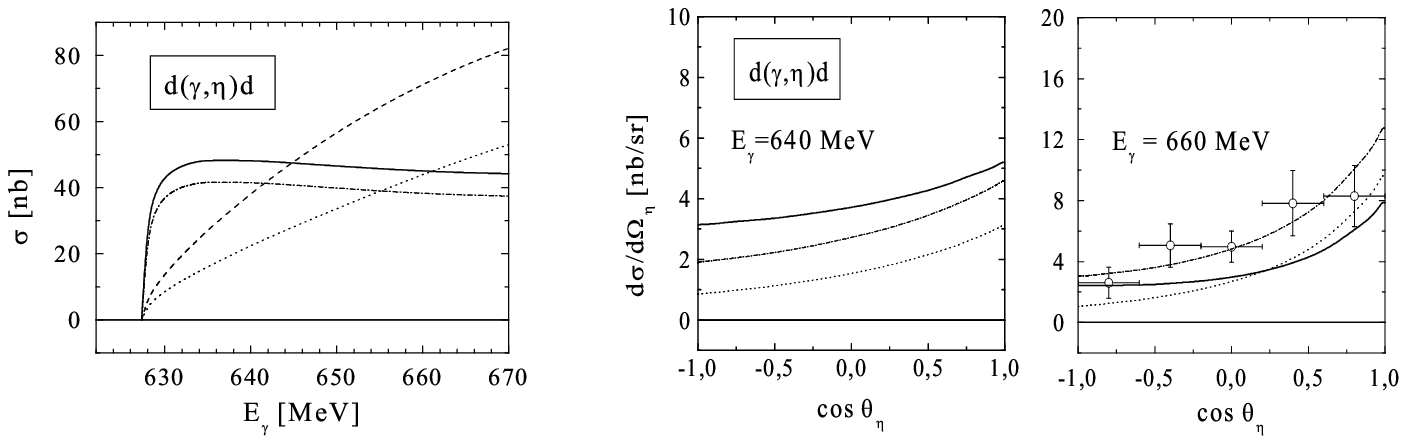,width=12cm}}
\vspace*{-.3cm}
\caption{Results for $d(\gamma,\eta)d$ from~\protect\cite{FiA02}. Left
panel: Total cross section. Middle and right panels: differential
cross sections. Dotted: IA; dashed: first order rescattering;
solid: complete 3-body model; dash-dot: 3-body model without
$\pi$-exchange contribution; Exp.: P.\ Hoffmann-Rothe {\it et
al.}, \Journal{\PRL}{78}{4697}{1997}.
}
\label{fig7a}
\vspace*{.1cm}
\centerline{\epsfig{file=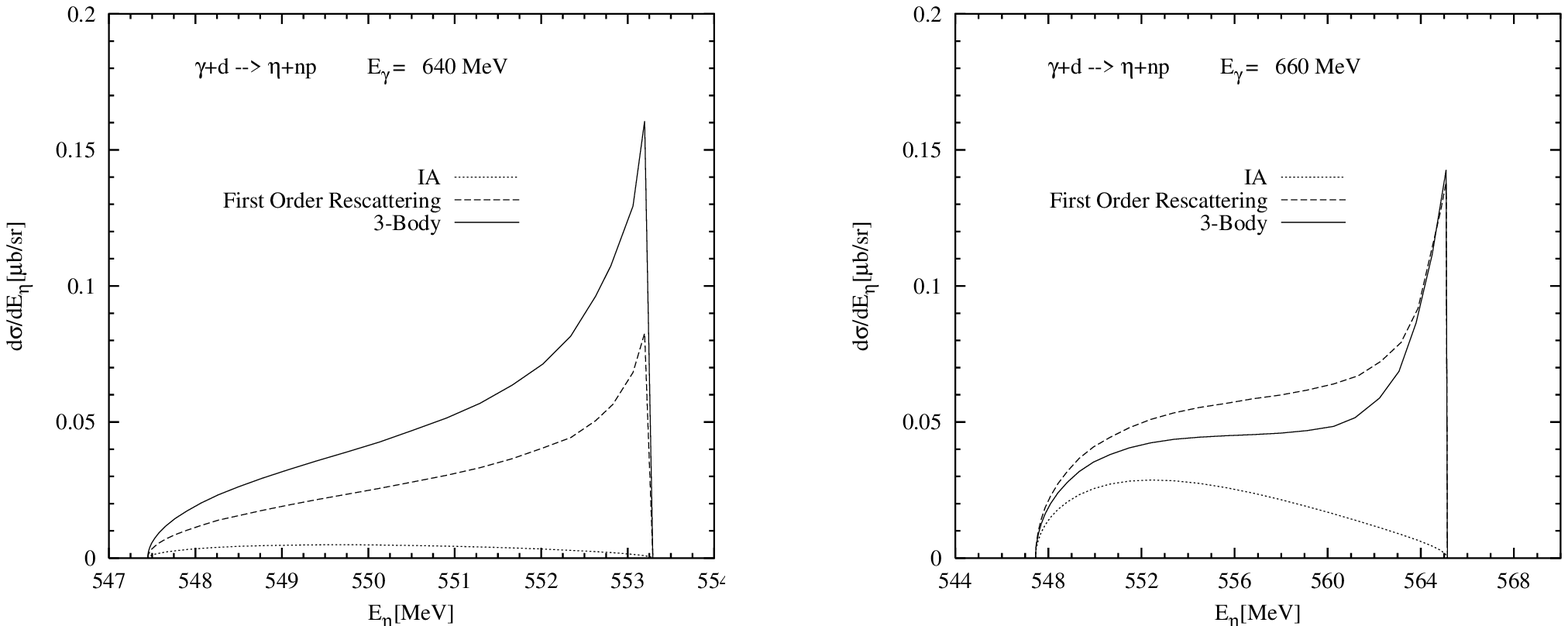,width=8cm}}
\vspace*{-.2cm}
\caption{$\eta$ meson spectrum for $d(\gamma,\eta)np$ close to
threshold from~\protect\cite{FiA02}.} 
\label{fig7}
\vspace*{-.5cm}
\end{figure}
Results for total and differential cross sections of the
coherent photoproduction are shown in Fig.~\ref{fig7a}. One readily
notes a drastic increase of the total cross section over the mere IA
right above threshold due to a strong attraction in the $(S=1,T=0)$
channel, which the first-order rescattering calculation is not able to
reproduce. Also in the differential cross sections in Fig.~\ref{fig7a}
a large difference between first-order calculation and complete
three-body approach is seen. 
\begin{figure}
\centerline {\epsfig{file=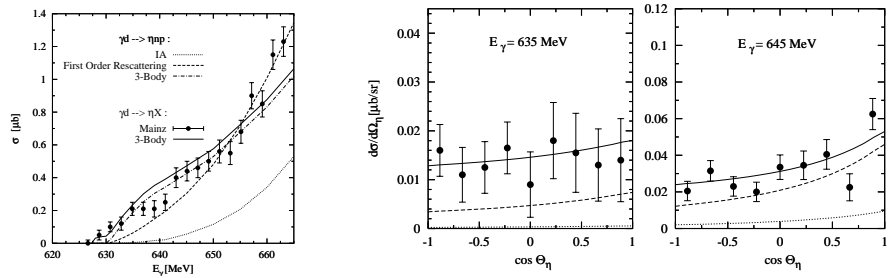,width=12cm}}
\vspace*{-.4cm}
\caption{Results for $d(\gamma,\eta)np$ from~\protect\cite{FiA02}. 
Inclusive data from Hejny {\it et al.}, \Journal{\EPJA}{13}{493}{2002}.
Left panel: Total cross section: 
(a) incoherent: dotted: IA; dashed: first order rescattering;
dash-dot: complete 3-body model; (b) inclusive: solid.
Middel and right panels: Differential cross sections: dotted: IA; 
dashed: first order rescattering; solid: complete 3-body model.
} 
\label{fig8}
\vspace*{-.6cm}
\end{figure}
Similar conclusions are reached for the
incoherent reaction. The importance of a three-body
treatment is demonstrated by the $\eta$ meson spectrum in
Fig.~\ref{fig7}. Close to threshold, the first order rescattering
underestimates significantly the three-body approach, whereas at
higher energies it yields an overestimation. 
Total and differential cross sections are shown in Fig.~\ref{fig8}.
The inclusive total cross section data exhibit a distinct enhancement
near threshold which is reproduced by the 3-body approach (left panel
of Fig.~\ref{fig8}). This is also the case for
the differential cross sections (middel and right panels of
Fig.~\ref{fig8}). It remains to be seen, whether a more realistic
calculation is also able to describe the data. 
\vspace*{-.3cm}

\section{Conclusions and Outlook} 
The main conclusions are: (i)
Near threshold only a three-body approach gives an adequate 
description of the $\eta NN$ dynamics.
The first order rescattering approximation for the final state
interaction fails drastically. (ii)
For $\eta d$ elastic scattering a very strong enhancement near 
threshold is found in the three-body calculation which is not born 
out in first order rescattering. The coupling to the $\pi NN$ channel
is relatively unimportant. (iii) 
With a {\sc Yamaguchi}-type separable interaction a satisfactory 
description of experimental data on $\eta$ photoproduction on the
deuteron is achieved, in particular the enhancement of the total cross
section right above the threshold, and the 
nearly isotropic angular distribution of the outgoing $\eta$ 
meson is reproduced in the three-body approach in contrast to the
first order rescattering.
(iv) The $\eta N$-interaction can be studied in incoherent eta 
production near threshold. However, a first order rescattering 
calculation as used, e.g. by A. Sibirtsev {\it et al.}, 
\Journal{\PRC}{65}{044007}{2002}, is not reliable for that purpose, 
because right above threshold a three-body approach is mandatory.

As open problems remain (i) the inclusion of additional two-body effects
like meson exchange curents as discussed in the coherent
process~\cite{RiA01}, and (ii) the use of realistic $NN$ interactions 
with inclusion of the tensor force and the deuteron $D$-wave.

\end{document}